\documentclass[12pt]{article}

\usepackage{amsfonts}
\usepackage{amsmath}
\usepackage{epsfig}
\usepackage{latexsym}
\usepackage{graphicx}
\usepackage{amssymb}
\usepackage{color}
\numberwithin{equation}{section}
\usepackage{url}
\allowdisplaybreaks

\def\spa#1{\phantom{\fbox{\rule[-#1cm]{0cm}{0cm}}}}

\def\be{\begin{equation}}
\def\ee{\end{equation}}
\def\bea{\begin{eqnarray}}
\def\eea{\end{eqnarray}}
\def\bequ{\begin{equation}}
\def\eequ{\end{equation}}

\renewcommand{\thefootnote}{\fnsymbol{footnote}}

\newcommand{\eq} {equation}

\begin{document}

\hfuzz=100pt
\title{$\mathcal{N}=2$ minimal model from 4d supersymmetric theory}
\date{}
\author{Masazumi Honda\footnote{masazumihondaAThri.res.in}
  \spa{0.5} \\
\\
{\small{\it Harish-Chandra Research Institute,}}
\\ {\small{\it Chhatnag Road, Jhusi, Allahabad 211019, India}} \\
}
\date{\small{August 2015}}

\maketitle
\thispagestyle{empty}
\centerline{}

\begin{abstract}
Previous studies have shown that
supersymmetric partition function on $T^2 \times S^2$ is related to elliptic genus of
two dimensional supersymmetric theory.
In this short note 
we find a four dimensional supersymmetric theory,
whose partition function on $T^2 \times S^2$ is
the same as elliptic genera of $\mathcal{N}=2$ minimal models
in two dimensions.
\end{abstract}
\vfill
\noindent HRI/ST/1510 

\renewcommand{\thefootnote}{\arabic{footnote}}
\setcounter{footnote}{0}

\newpage
\setcounter{page}{1}
\section{Introduction}
Conformal field theories (CFT) in higher dimensions are usually 
less under control than two dimensional CFTs.
This is mainly because higher dimensional conformal symmetry is finite dimensional in contrast to 2d conformal symmetry.
It would be illuminating
if higher dimensional CFTs have some relations to 2d CFTs in certain ways.

Recently supersymmetric partition function on $T^2 \times S^2$ has provided
some interesting aspects along this direction.
Imposing appropriate boundary conditions,
this partition function can be expressed as 
the supersymmetric index \cite{Closset:2013sxa}
\begin{\eq}
Z_{T^2 \times S^2} = {\rm Tr}\Bigl[ (-1)^F q^P x^{J_3} \prod_a t_a^{F_a} \Bigr] ,
\label{eq:4dindex}
\end{\eq}
where $F $ is Fermion number, $P $ is spatial momentum of $T^2 $, 
$J^3 $ is angular momentum on $S^2$ 
and $F_a$ is flavor charge.
Given this index formula,
one may recall the elliptic genus
\begin{\eq}
Z_{T^2} 
= {\rm Tr }_{\rm R}
\Bigl[ (-1)^F   q^{H_L} \bar{q}^{H_R }  \prod_a t_a^{F_a} \Bigr] 
= {\rm Tr }_{\rm R}
\Bigl[ (-1)^F q^P  \prod_a t_a^{F_a} \Bigr] ,
\end{\eq}
which is equivalent to supersymmetric partition function on $T^2$. 
In fact previous works have found that
partition function of 4d $\mathcal{N}=1$ supersymmetric theory on $T^2 \times S^2$
have simple relations to elliptic genera of 2d $\mathcal{N}=(0,2)$
supersymmetric theories \cite{Closset:2013sxa,Nishioka:2014zpa,Benini:2015noa,Honda:2015yha} (see also \cite{Closset:2015rna}).
This fact provides
nontrivial relations between properties of 4d and 2d supersymmetric theories.
One of interesting directions is 
to pursue relations between 4d and 2d dualities \cite{Honda:2015yha,OsakaLec,Gadde:2015wta,Putrov:2015jpa}.
For example one can show \cite{Honda:2015yha,OsakaLec} that
two dimensional $(0,2)$ triality \cite{Gadde:2013lxa} for elliptic genus
comes from 4d Seiberg-like duality \cite{Seiberg:1994pq} for partition function on $T^2 \times S^2$.

In this short note
we discuss that
certain 4d supersymmetric theory is related to $\mathcal{N}=2$ minimal models of ADE type
in two dimensions.
More concretely
we study partition function on $T^2 \times S^2$ of this 4d theory and
show that this is the same as elliptic genera of the $\mathcal{N}=2$ minimal models, 
which have been studied well in \cite{Witten:1993jg,DiFrancesco:1993ty,Kawai:1993jk}.
In the rest of this note,
we first briefly review
some aspects of 4d $\mathcal{N}=1$ supersymmetric theory on $T^2 \times S^2$ and
explain exact formula for the $T^2 \times S^2$ partition function 
in section \ref{sec:review}.
Then we find the four dimensional supersymmetric theory,
whose partition function on $T^2 \times S^2$ 
agrees with the elliptic genera of the $\mathcal{N}=2$ minimal models 
in section \ref{sec:construction}.

\section{Supersymmetric partition function on $T^2 \times S^2$}
\label{sec:review}
First we briefly review 
some relevant facts on 4d $\mathcal{N}=1$ supersymmetric theory on $T^2 \times S^2$
and write down exact formula for the $T^2 \times S^2$ partition function
obtained in the previous studies \cite{Closset:2013sxa,Nishioka:2014zpa,Benini:2015noa,Honda:2015yha}.

We define $T^2 \times S^2$ as a quotient of $\mathbb{C}\times S^2$.
Denoting complex coordinates of $\mathbb{C}$ and $S^2$ by $w$ and $z$, respectively,
we regard $T^2 \times S^2$ as the following identification of $\mathbb{C}\times S^2$: 
\begin{\eq}
(w ,z )  \sim  (w  +2\pi ,  e^{2\pi i\alpha }z) 
\sim ( w   +2\pi \tau ,   e^{2\pi i \beta}z   ) ,
\label{eq:space}
\end{\eq}
where $\tau $ is the complex structure parameter of $T^2$ 
and $(\alpha$, $\beta )$ are real parameters satisfying $\alpha \sim \alpha +1$, $\beta \sim \beta +1$.
In order to preserve supersymmetry on this background,
the Killing spinors $\zeta ,\tilde{\zeta}$ have to satisfy
\begin{\eq}
(\nabla_\mu -iA_\mu )\zeta =0 ,\quad (\nabla_\mu +iA_\mu )\tilde{\zeta} =0 ,
\end{\eq}
where $A_\mu$ is the background gauge field of R-symmetry 
with the monopole configuration\footnote{
We are using a freedom to take real $A$.
This notation corresponds to $s=1$ and $\kappa =0$ in \cite{Assel:2014paa}.
}
\begin{\eq}
A= -\frac{i}{2} \frac{\bar{z}dz -zd\bar{z}}{1+z\bar{z}} .
\end{\eq}
This indicates that 
fields on $T^2 \times S^2$ generically have magnetic charges 
depending on their R-charges \cite{Closset:2013vra,Closset:2013sxa}
since the R-symmetry background gauge field of the monopole configuration appears in every covariant derivative of the fields.
More generally we can also turn on magnetic background gauge field of flavor symmetry with integer flux $g$.
Hence if we have the magnetic flux $g$ of the flavor symmetry,
then a chiral multiplet with the R-charge $r$ and flavor charge $q_f$
have the magnetic charge
\begin{\eq}
\mathbf{r} = r+q_f g ,
\label{eq:magnetic_charge}
\end{\eq}
which should be integer in order to
satisfy quantization condition for the magnetic flux on $S^2$.

Now we explicitly write down 
the exact formula for the $T^2 \times S^2$ partition function of 
supersymmetric theory with only chiral multiplets.
Taking appropriate boundary conditions,
this partition function can be represented as the supersymmetric index \cite{Closset:2013sxa}
\[
Z_{T^2 \times S^2} = {\rm Tr}\Bigl[ (-1)^F q^P x^{J_3} \prod_a t_a^{F_a} \Bigr] .
\]
The parameters $q,x$ and $t_a$ are defined by
\begin{\eq}
q=e^{2\pi i\tau},\quad 
x=e^{2\pi i \sigma},\quad t_a = e^{2\pi i\xi_a} ,
\label{eq:fugacity}
\end{\eq}
where $(\tau ,\sigma =\tau\alpha  -\beta )$ is complex structures of $T^2 \times S^2 $
and $\xi_a$ is fugacity of flavor symmetry.
For 4d $\mathcal{N}=1$ theory with only chiral multiplets,
the partition function takes the following form \cite{Closset:2013sxa,Nishioka:2014zpa,Benini:2015noa,Honda:2015yha}
\begin{\eq}
Z_{T^2 \times S^2} = \prod_I Z^{(\mathbf{r}_I )} (\tau ,\sigma ,\xi_a ) ,
\end{\eq}
where $Z^{(\mathbf{r}_I )}$ is the contribution from a 4d $\mathcal{N}=1$ chiral multiplet
with magnetic charge $\mathbf{r}_i$:
\begin{\eq}
Z^{(\mathbf{r} )} (\tau , \sigma ,\xi_a )
 =\left\{ \begin{matrix}
 \prod_{m=-\frac{\mathbf{r}}{2}+1}^{\frac{\mathbf{r}}{2}-1} \frac{i\theta_1 (\tau | m\sigma  +\sum_a q_f^{a} \xi_a )}{\eta (\tau )}
& {\rm for}\ \mathbf{r}>1  \cr
1& {\rm for}\ \mathbf{r}=1 \cr
 \prod_{m=-\frac{|\mathbf{r}|}{2}}^{\frac{|\mathbf{r}|}{2}}  \frac{i\eta (\tau )}{\theta_1 (\tau |m\sigma  +\sum_a q_f^{a} \xi_a )} .
& {\rm for}\ \mathbf{r}<1
\end{matrix}\right. .
\label{eq:det}
\end{\eq}
Here the Dedekind eta function $\eta (\tau )$ and 
Jacobi theta function $\theta_1 (\tau |z)$ are defined by
\begin{\eq}
\eta (\tau ) = q^{\frac{1}{24}} \prod_{n=1}^\infty (1-q^n ) ,\quad
\theta_1 (\tau |z)
=-iq^{\frac{1}{8}}y^{\frac{1}{2}} \prod_{k=1}^\infty (1-q^k )(1-yq^k ) (1-y^{-1}q^{k-1}) ,
\end{\eq}
with $y=e^{2\pi iz}$.
Note that the first and third lines in the RHS of \eqref{eq:det} are formally the same\footnote{
In the 2d picture, the fugacity $\sigma$ can be regarded as
the one of a flavor symmetry in 2d.
} as contributions to elliptic genus from 2d $\mathcal{N}=(0,2)$ Fermi and chiral multiplets, 
respectively \cite{Benini:2013nda,Gadde:2013dda,Benini:2013xpa}.

\section{$\mathcal{N}=2$ minimal model from 4d theory on $T^2 \times S^2$}
\label{sec:construction}
\begin{table}[t]
\begin{center}
  \begin{tabular}{|c|c   c c |  }
  \hline              & $U(1)_R$ & $U(1)_f$ & $U(1)_Q$   \\
\hline  $\Phi$    &    1         &    -1      &   0      \\
    $\tilde{\Phi}$  &    1         &   1        &   0     \\
           $Q_I$       & 1/2   &  -1/2     &  $w_I +1/2$   \\  
      $\tilde{Q}_I$  & 1/2   & 3/2       &  $-w_I -1/2$    \\   \hline
         fugacity   &   -      &     $z$        &  $z$    \\  
 magnetic flux   &     1    &     1            &   0   \\\hline
  \end{tabular}
\end{center}
\caption{Field content of the 4d supersymmetric theory.
}
\label{tab:content}
\end{table}

Now we are ready to construct the 4d supersymmetric theory,
whose $T^2 \times S^2$ partition function is
the same as the elliptic genera of the $\mathcal{N}=2$ minimal models.
At first sight
one might think that this could be easily done by engineering Landau-Ginzburg models from 4d,
which are known to correspond to the $\mathcal{N}=2$ minimal models \cite{Witten:1993jg,DiFrancesco:1993ty,Kawai:1993jk}.
However, this is impossible for most cases 
since chiral multiplets must have integer magnetic charges on $T^2 \times S^2$.
Let us consider the 4d theory with the field content of table \ref{tab:content} and the superpotential
\begin{\eq}
W =\sum_{I=1}^N \tilde{Q}_I \Phi Q_I +\tilde{\Phi}\Phi .
\end{\eq}
Then the $T^2 \times S^2$ partition function of this theory is represented as
\begin{\eq}
Z_{\rm T^2 \times S^2}
= {\rm Tr}\Bigl[ (-1)^F q^P x^{J_3}t_f^{F_f} t_Q^{F_Q} \Bigr] .
\end{\eq}
The parameters $t_f = e^{2\pi i\xi_f}$ and $t_Q = e^{2\pi i\xi_Q}$ are
the fugacities of the $U(1)_f$ and $U(1)_Q$ flavor symmetries respectively
but here we identify these two fugacities, namely $\xi_f =\xi_Q =z$.

Now we write down the partition function by using the results in \cite{Closset:2013sxa,Nishioka:2014zpa,Benini:2015noa,Honda:2015yha}.
Contributions from each chiral multiplet are given by
\begin{\eq}
Z_\Phi = \frac{i\eta (q)}{\theta_1 (\tau |-z)} ,\quad
Z_{\tilde{\Phi}} = \frac{i\theta_1 (\tau |z)}{\eta (q)} ,\quad
Z_{Q_I} = \frac{i\eta (q)}{\theta_1 (\tau | w_I z)} ,\quad
Z_{\tilde{Q}_I} = \frac{i\theta_1 (\tau |(1-w_I )z)}{\eta (q)}  .
\end{\eq}
Combining these expressions, we obtain\footnote{Note $\theta_1 (\tau |-z) =-\theta_1 (\tau |z)$.}
\begin{\eq}
Z_{T^2 \times S^2} =Z_\Phi Z_{\tilde{\Phi}} \prod_{I=1}^N Z_{Q_I} Z_{\tilde{Q}_I}
= \prod_{I=1}^N \frac{\theta_1 (\tau |(1-w_I )z)}{\theta_1 (\tau | w_I z)} ,
\end{\eq}
up to overall signature.
This expression is exactly the same 
as the elliptic genera of the $\mathcal{N}=2$ minimal models \cite{Witten:1993jg,DiFrancesco:1993ty,Kawai:1993jk}
associated with the ADE singularity $W=0$,
where $W(x_1 ,\cdots ,x_N )$ is the polynomial of $x$ satisfying
\begin{\eq}
W(\lambda^{w_1}x_1 ,\cdots ,\lambda^{w_N}x_N ) =\lambda W(x_1 ,\cdots ,x_N ) .
\end{\eq}
Unfortunately
we have not found why the $\mathcal{N}=2$ minimal models appear
from the 4d supersymmetric theory physically.
One might think that this result was similar to AGT relation \cite{Alday:2009aq}
but this does not seem true since the 4d theory here is is not gauge theory.
It is interesting to find any physical interpretations.

In this note we have constructed 4d supersymmetric theories,
whose partition function on $T^2 \times S^2$
matches with the elliptic genera of the $\mathcal{N}=2$ ADE minimal models.
Our result and the previous results \cite{Honda:2015yha,OsakaLec,Gadde:2015wta,Putrov:2015jpa} would imply that
sub-sector of 4d $\mathcal{N}=1$ supersymmetric theory have hidden infinite dimensional symmetry at infrared fixed point.
It is interesting 
if one relates the results to recent discussions on hidden symmetries 
in 4d theories \cite{Beem:2013sza,Strominger:2013lka,He:2015zea}.
One of more challenging directions is 
to consider non-supersymmetric gauge theory.
It has been expected that
confining gauge theories in planar limit have descriptions 
in terms of weakly coupled string theories \cite{'tHooft:1973jz,Witten:1979kh}.
This implies that
the gauge theories in confined phase have 2d CFT descriptions.
Recently there appears a concrete proposal regarding this for 4d pure $SU(N)$ Yang-Mills theory on $S^1 \times S^3$ 
in the planar limit \cite{Basar:2015xda}.
It would be attractive to consider non-supersymmetric theories on $T^2 \times S^2$ in confined phase.

\subsection*{Acknowledgment}
This work is motivated by the previous collaboration \cite{Honda:2015yha} with Yutaka Yoshida.

\providecommand{\href}[2]{#2}\begingroup\raggedright\endgroup


\begin{thebibliography}{10}

\bibitem{Closset:2013sxa}
C.~Closset and I.~Shamir, {\it {The $\mathcal{N}=1$ Chiral Multiplet on
  $T^2\times S^2$ and Supersymmetric Localization}},  {\em JHEP} {\bf 1403}
  (2014) 040, [\href{http://arxiv.org/abs/1311.2430}{{\tt arXiv:1311.2430}}].

\bibitem{Nishioka:2014zpa}
T.~Nishioka and I.~Yaakov, {\it {Generalized indices for $ \mathcal{N} $ = 1
  theories in four-dimensions}},  {\em JHEP} {\bf 1412} (2014) 150,
  [\href{http://arxiv.org/abs/1407.8520}{{\tt arXiv:1407.8520}}].

\bibitem{Benini:2015noa}
F.~Benini and A.~Zaffaroni, {\it {A topologically twisted index for
  three-dimensional supersymmetric theories}},  {\em JHEP} {\bf 07} (2015) 127,
  [\href{http://arxiv.org/abs/1504.03698}{{\tt arXiv:1504.03698}}].

\bibitem{Honda:2015yha}
M.~Honda and Y.~Yoshida, {\it {Supersymmetric index on $T^2 x S^2$ and elliptic
  genus}},  \href{http://arxiv.org/abs/1504.04355}{{\tt arXiv:1504.04355}}.

\bibitem{Closset:2015rna}
C.~Closset, S.~Cremonesi, and D.~S. Park, {\it {The equivariant A-twist and
  gauged linear sigma models on the two-sphere}},  {\em JHEP} {\bf 06} (2015)
  076, [\href{http://arxiv.org/abs/1504.06308}{{\tt arXiv:1504.06308}}].

\bibitem{OsakaLec}
Y.~Tachikawa {\em Intensive lecture at Osaka University} (October, 2014).

\bibitem{Gadde:2015wta}
A.~Gadde, S.~S. Razamat, and B.~Willett, {\it {On the reduction of $4d$
  $\mathcal{N}=1$ theories on $\mathbb {S}^2$}},
  \href{http://arxiv.org/abs/1506.08795}{{\tt arXiv:1506.08795}}.

\bibitem{Putrov:2015jpa}
P.~Putrov, J.~Song, and W.~Yan, {\it {$(0, 4)$ dualities}},
  \href{http://arxiv.org/abs/1505.07110}{{\tt arXiv:1505.07110}}.

\bibitem{Gadde:2013lxa}
A.~Gadde, S.~Gukov, and P.~Putrov, {\it {(0, 2) trialities}},  {\em JHEP} {\bf
  1403} (2014) 076, [\href{http://arxiv.org/abs/1310.0818}{{\tt
  arXiv:1310.0818}}].

\bibitem{Seiberg:1994pq}
N.~Seiberg, {\it {Electric - magnetic duality in supersymmetric nonAbelian
  gauge theories}},  {\em Nucl.Phys.} {\bf B435} (1995) 129--146,
  [\href{http://arxiv.org/abs/hep-th/9411149}{{\tt hep-th/9411149}}].

\bibitem{Witten:1993jg}
E.~Witten, {\it {On the Landau-Ginzburg description of N=2 minimal models}},
  {\em Int. J. Mod. Phys.} {\bf A9} (1994) 4783--4800,
  [\href{http://arxiv.org/abs/hep-th/9304026}{{\tt hep-th/9304026}}].

\bibitem{DiFrancesco:1993ty}
P.~Di~Francesco, O.~Aharony, and S.~Yankielowicz, {\it {Elliptic genera and the
  Landau-Ginzburg approach to N=2 orbifolds}},  {\em Nucl. Phys.} {\bf B411}
  (1994) 584--608, [\href{http://arxiv.org/abs/hep-th/9306157}{{\tt
  hep-th/9306157}}].

\bibitem{Kawai:1993jk}
T.~Kawai, Y.~Yamada, and S.-K. Yang, {\it {Elliptic genera and N=2
  superconformal field theory}},  {\em Nucl. Phys.} {\bf B414} (1994) 191--212,
  [\href{http://arxiv.org/abs/hep-th/9306096}{{\tt hep-th/9306096}}].

\bibitem{Assel:2014paa}
B.~Assel, D.~Cassani, and D.~Martelli, {\it {Localization on Hopf surfaces}},
  {\em JHEP} {\bf 1408} (2014) 123, [\href{http://arxiv.org/abs/1405.5144}{{\tt
  arXiv:1405.5144}}].

\bibitem{Closset:2013vra}
C.~Closset, T.~T. Dumitrescu, G.~Festuccia, and Z.~Komargodski, {\it {The
  Geometry of Supersymmetric Partition Functions}},  {\em JHEP} {\bf 1401}
  (2014) 124, [\href{http://arxiv.org/abs/1309.5876}{{\tt arXiv:1309.5876}}].

\bibitem{Benini:2013nda}
F.~Benini, R.~Eager, K.~Hori, and Y.~Tachikawa, {\it {Elliptic genera of
  two-dimensional N=2 gauge theories with rank-one gauge groups}},  {\em
  Lett.Math.Phys.} {\bf 104} (2014) 465--493,
  [\href{http://arxiv.org/abs/1305.0533}{{\tt arXiv:1305.0533}}].

\bibitem{Gadde:2013dda}
A.~Gadde and S.~Gukov, {\it {2d Index and Surface operators}},  {\em JHEP} {\bf
  1403} (2014) 080, [\href{http://arxiv.org/abs/1305.0266}{{\tt
  arXiv:1305.0266}}].

\bibitem{Benini:2013xpa}
F.~Benini, R.~Eager, K.~Hori, and Y.~Tachikawa, {\it {Elliptic Genera of 2d
  ${\mathcal{N}}$ = 2 Gauge Theories}},  {\em Commun. Math. Phys.} {\bf 333}
  (2015), no.~3 1241--1286, [\href{http://arxiv.org/abs/1308.4896}{{\tt
  arXiv:1308.4896}}].

\bibitem{Alday:2009aq}
L.~F. Alday, D.~Gaiotto, and Y.~Tachikawa, {\it {Liouville Correlation
  Functions from Four-dimensional Gauge Theories}},  {\em Lett.Math.Phys.} {\bf
  91} (2010) 167--197, [\href{http://arxiv.org/abs/0906.3219}{{\tt
  arXiv:0906.3219}}].

\bibitem{Beem:2013sza}
C.~Beem, M.~Lemos, P.~Liendo, W.~Peelaers, L.~Rastelli, et~al., {\it {Infinite
  Chiral Symmetry in Four Dimensions}},  {\em Commun.Math.Phys.} {\bf 336}
  (2015), no.~3 1359--1433, [\href{http://arxiv.org/abs/1312.5344}{{\tt
  arXiv:1312.5344}}].

\bibitem{Strominger:2013lka}
A.~Strominger, {\it {Asymptotic Symmetries of Yang-Mills Theory}},  {\em JHEP}
  {\bf 1407} (2014) 151, [\href{http://arxiv.org/abs/1308.0589}{{\tt
  arXiv:1308.0589}}].

\bibitem{He:2015zea}
T.~He, P.~Mitra, and A.~Strominger, {\it {2D Kac-Moody Symmetry of 4D
  Yang-Mills Theory}},  \href{http://arxiv.org/abs/1503.02663}{{\tt
  arXiv:1503.02663}}.

\bibitem{'tHooft:1973jz}
G.~'t~Hooft, {\it {A Planar Diagram Theory for Strong Interactions}},  {\em
  Nucl. Phys.} {\bf B72} (1974) 461.

\bibitem{Witten:1979kh}
E.~Witten, {\it {Baryons in the 1/n Expansion}},  {\em Nucl. Phys.} {\bf B160}
  (1979) 57.

\bibitem{Basar:2015xda}
G.~Basar, A.~Cherman, K.~R. Dienes, and D.~A. McGady, {\it {A 4D-2D equivalence
  for large-N Yang-Mills theory}},  \href{http://arxiv.org/abs/1507.08666}{{\tt
  arXiv:1507.08666}}.

\end{thebibliography}
\end{document}